\documentclass[prl,twocolumn,showpacs,superscriptaddress,floats]{revtex4-1}

\usepackage{graphicx}

\begin{document}

\title{The momentum space electronic structure and the charge order of high temperature Ca$_{2-x}$Na$_{x}$CuO$_2$Cl$_2$
and Bi$_2$Sr$_2$CaCu$_2$O$_{8+\delta}$ superconductors}

\author{Jian-Qiao Meng}
\affiliation{Department of Physics, University of California, Santa
Cruz, CA 95064, USA}

\author{M.~Brunner}
\affiliation{Department of Physics, University of California, Santa
Cruz, CA 95064, USA}

\author{K.-H.~Kim}
\affiliation{Department of Physics, Pohang University of Science and Technology, Pohang 790784, South Korea}

\author{H.-G.~Lee}
\affiliation{Department of Physics, Pohang University of Science and Technology, Pohang 790784, South Korea}

\author{S.-I.~Lee}
\affiliation{Department of Physics, Pohang University of Science and Technology, Pohang 790784, South Korea}

\author{J.~S.~Wen}
\affiliation{Condensed Matter Physics and Materials Science
Department, Brookhaven National Laboratory, Upton, New York 11973,
USA}

\author{Z.~J.~Xu}
\affiliation{Condensed Matter Physics and Materials Science
Department, Brookhaven National Laboratory, Upton, New York 11973,
USA}

\author{G.~D.~Gu}
\affiliation{Condensed Matter Physics and Materials Science
Department, Brookhaven National Laboratory, Upton, New York 11973,
USA}

\author{G.-H.~Gweon}
\email{Corresponding author: gweon@ucsc.edu} \affiliation{Department
of Physics, University of California, Santa Cruz, CA 95064, USA}

\date{\today}

\begin{abstract}

We study the electronic structure of Ca$_{2-x}$Na$_{x}$CuO$_2$Cl$_2$ and
Bi$_2$Sr$_2$CaCu$_2$O$_{8+\delta}$ samples in a wide range of doping, using
angle-resolved photoemission spectroscopy, with emphasis on the Fermi
surface (FS) in the near anti-nodal region.  The ``nesting wave vector,''
i.e., the wave vector that connects two nearly flat pieces of the FS in the
anti-nodal region, reveals a universal monotonic decrease in magnitude as a
function of doping.  Comparing our results to the charge order recently
observed by scanning tunneling spectroscopy (STS), we conclude that the FS
nesting and the charge order pattern seen in STS do not have a direct
relationship.  Therefore, the charge order likely arises due to strong
correlation physics rather than FS nesting physics.

\end{abstract}

 \maketitle

The origin of the ``pseudo-gap'' remains a central mystery in the
physics of high-temperature superconductors. The pseudo-gap
\cite{Ding1996,Timusk1999} means that the single-particle spectral
function $A(\vec{k},\omega)$ shows depleted spectral weight at the
Fermi energy ($E_F$) in the metallic phase above the superconducting
transition temperature, in stark contrast to a Landau quasi-particle
peak at $E_F$ that is expected for a normal Fermi liquid. By now, the
pseudo-gap is routinely observed by experimental tools that probe
$A(\vec{k},\omega)$, the angle-resolved photoelectron spectroscopy
(ARPES) or, its Fourier transform, the scanning tunneling
spectroscopy (STS).

Recently, quite a few ARPES and STS studies \cite{KTanaka2006,
TKondo2007, WSLee2007, KTerashima2007, TKondo2009, Boyer2007,
Kohsaka2008, Hanaguri2007} have contributed to an emerging
phenomenology of the pseudo-gap.  In this emerging picture, the
pseudo-gap is separate from the superconducting gap, is dominant in
the ``anti-nodal'' region in the momentum space, and is
characterized by a ``checkerboard pattern'' of charge order in the real
space.  In particular, this checkerboard pattern, representing an
organization of the electron density with approximately four lattice constant
periodicities along the $a$ or $b$ direction of the (nearly) tetragonal
CuO$_2$ lattice, is observed in all the samples of cuprates that have been
found to be appropriate for STS studies, i.e., \@
Bi$_2$Sr$_2$CaCu$_2$O$_{8+\delta}$ (Bi2212) compounds
\cite{Hoffman2002A,McElroy2005,Howald2003A,Howald2003B, YHLiu2007,
Vershinin2004}, Bi$_{2-y}$Pb$_y$Sr$_{2-z}$La$_z$CuO$_{6+\delta}$ (Bi2201)
compounds \cite{Wise2008}, and Ca$_{2-x}$Na$_{x}$CuO$_2$Cl$_2$ (Na-CCOC)
compounds \cite{Hanaguri2004,Hanaguri2007}.  This unique feature, while strong
in the underdoped regime, is observed up to near-optimal doping
\cite{McElroy2005} or optimal doping \cite{Hanaguri2007}, albeit with a
diminished weight.

What is the underlying mechanism of the checkerboard pattern?  Here
we shall differentiate between two generic scenarios, a weak-correlation
scenario and a strong-correlation scenario.  In the first
scenario, we associate the checkerboard pattern as reflecting the
instability of the electron system due to a Fermi-surface (FS)
nesting.  Thus, here we are envisioning the standard Peierls-type
charge-density-wave (CDW) phenomenon \cite{Peierls} or the standard
weak-coupling spin-density-wave (SDW) phenomenon \cite{Gruner}.  In
the second scenario, we consider the checkerboard pattern as driven
by large-energy scale physics, as opposed to the FS nesting physics.
For instance, the checkerboard pattern has been attributed to an
instability of the Hubbard model \cite{Seibold2007}, or to the formation of
a Wigner solid of hole pairs embedded in a sea of $d$-wave resonating
valence bond (RVB) states \cite{Anderson2004}.  The essential criterion by
which we can distinguish the two mechanisms is the agreement between the
Fermi-surface nesting wave vector and the checkerboard periodicity.
Within the second scenario, this agreement is not a required primary
feature but an optional feature. Within the first scenario, this agreement
is an absolute requirement.

In our definition above, we take the weak-correlation FS-based scenario
as the FS nesting scenario only, leaving out other, generally possible,
one-electron-band mechanisms such as the van Hove singularity-induced CDW
\cite{Rice-CDW}.  Note also that a FS object such as the
``Fermi arc'' \cite{Norman1998} may give rise to a high Lindhard
susceptibility due to the high density of states
at extremal points of the arc, causing a CDW\@.
Such a Fermi-arc-based CDW scenario, e.g.,\@ as
suggested in a recent STS work \cite{Kohsaka2008}, requires a fundamental
modification of the FS to a Fermi arc with a strong correlation as a {\em prerequisite}
and does {\em not} involve FS nesting.  Thus, such a scenario should be
considered as a strong-correlation scenario, not as a weak correlation
scenario.

\begin{figure*}[floatfix]
\begin{center}
\includegraphics[width=1.82\columnwidth,angle=0]{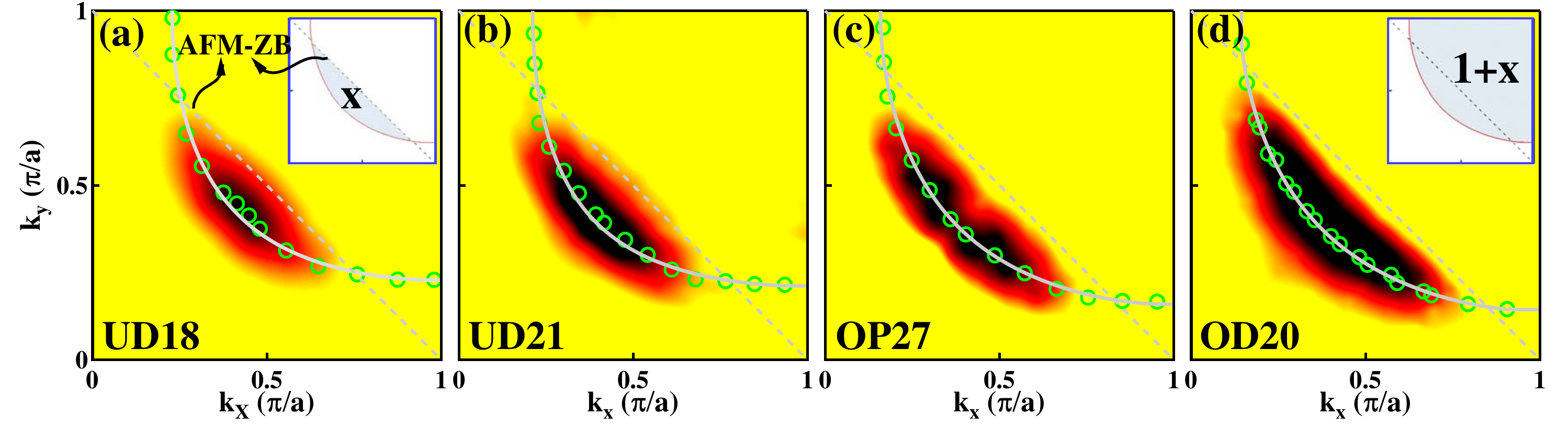}
\end{center}
\caption{(Color online) Doping evolution of the Fermi energy intensity maps for (a) UD18,
(b) UD21, (c) OP27, and (d) OD20 Na-CCOC samples.  The energy integration
window [-10 meV, +10 meV] centered at $E_F$ was used. These maps are
obtained by symmetrizing the original data with respect to the
(0,0)--($\pi$,$\pi$) line.  The Fermi momentum values determined by the
maximum of the MDC at $E_F$ are marked by open circles (green).  The solid
lines (gray) show fits to the Fermi surface of the tight-binding band theory \cite{Markiewicz2005}.  In the color map, dark (black) means
high intensity and pale (yellow) means low intensity.  The insets of (a), and (d) are discussed later(``Luttinger sum rule'').}
\end{figure*}

Thus, we pose the subsequent question: For high-temperature
superconductors, is there a direct connection between the
Fermi-surface nesting and the checkerboard pattern?  A survey of the
literature does not give a clear answer, which this paper now aims
to provide.  One does find in the literature, however, some initial
attempts at such an answer.  For example, a recent work on Bi2201
superconductors \cite{Wise2008} found that the periodicity of the
charge order pattern is a strong function of the doping, ranging
from 4.5 lattice constants in the underdoped region to 6.2 lattice
constants at the optimal doping.  Qualitatively, this trend is what
one would expect for a weak-correlation scenario, as the authors of
that work have indeed suggested. In addition, this conclusion
appears to be supported by work on Na-CCOC compounds, where a good
agreement between the Fermi-surface nesting vector in ARPES
\cite{KMShen2005} and the checkerboard periodicity
\cite{Hanaguri2004} has been noted. However, this study was limited
to low doping values, and in view of more recent work
\cite{Hanaguri2007}, a study covering a wider doping range including
optimal doping is necessary to attain a firm conclusion.

In this paper, we report ARPES data on Na-CCOC samples and Bi2212
samples in view of these questions.  We significantly widen the
doping range of the Na-CCOC samples studied by ARPES in comparison
to previous studies \cite{KMShen2005,Sasagawa2003}, and find that
such an improvement is of essential importance.  We compare our
results on Na-CCOC and Bi2212 samples with their known checkerboard
periodicity values.  We re-examine the degree of agreement, or
disagreement, between the Fermi-surface nesting and the checkerboard
pattern for Na-CCOC samples, and, in addition, Bi2201 samples. We
find that in all the cases examined, such an agreement is absent, or
accidental at best.  Thus, we find that a strong-correlation scenario
is much more likely than a weak-correlation scenario for explaining the
pseudo-gap.

\begin{figure}[b]
\setlength{\abovecaptionskip}{0pt}
\setlength{\belowcaptionskip}{0pt}
\begin{center}
\includegraphics[width=0.9\columnwidth,angle=0]{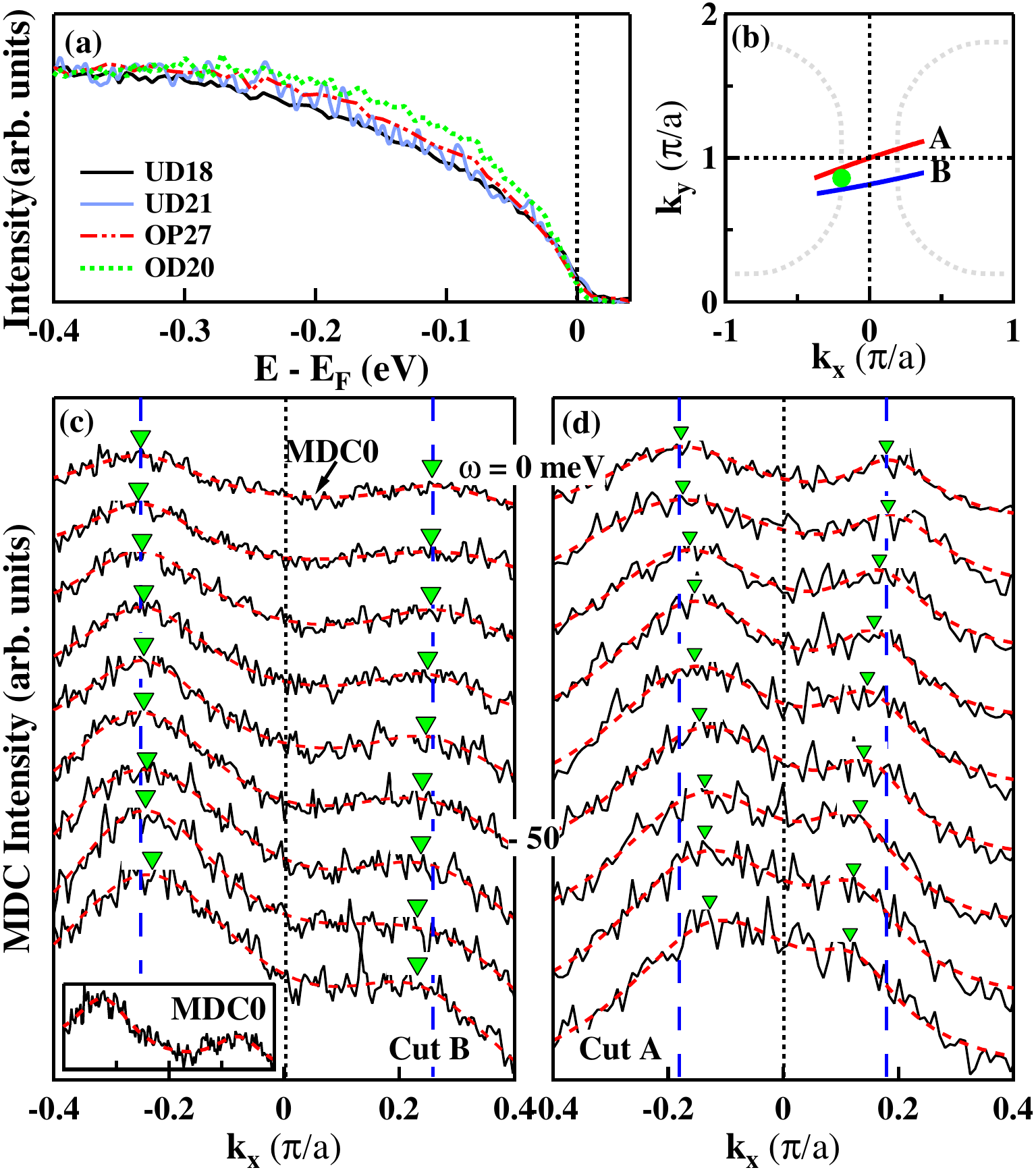}
\end{center}
\caption{(Color online) (a) EDCs in the anti-nodal region [see (b) for the $\vec{k}$ information], as a function of doping for Na-CCOC.  (b) $\vec{k}$ space cuts A,B corresponding to the data of (d), and (c), respectively, and the approximate $\vec{k}$ value (point) corresponding to the data of (a).
(c), and (d) MDCs as a function of the binding energy for UD21 (c) and OP27 (d).
Red dashed lines are fit curves (see text) and the vertical dashed lines (blue)
correspond to the peak positions at $\omega = 0$, i.e.\ at $E_F$.  The curves are separated
vertically for easy view, and the binding energy values for adjacent curves differ by 10 meV\@.
The inset of (c) re-displays the $E_F$ data, emphasizing the clear two peak structure.
The two peaks for each curve are marked by triangular symbols, each of which
indicates both the peak position and its uncertainty (horizontal size) as determined by the MDC fit procedure.}
\end{figure}

The photoemission measurements were carried out at beam line 5-4 of
the Stanford Synchrotron Radiation Lightsource (SSRL) and at beam
line 10.0.1 of the Advanced Light Source (ALS) at Lawrence Berkeley
National Laboratory, using a Scienta R4000 electron energy analyzer.
The photon energy used was 25.5 eV for Na-CCOC and 25 eV for Bi2212,
both with an energy resolution of 15-20 meV\@. The angular resolution is
$\sim$ 0.2$^{\circ}$ (0.008 {\AA}$^{-1}$ in momentum). Na-CCOC
samples, grown by a high-pressure flux method to unprecedented high
doping values for this experiment, include underdoped samples ($T_c
= 18, and 21$ K), optimally doped samples ($T_c = 27$ K), and overdoped
samples ($T_c = 20$ K) \cite{KyungHee2006}. Bi2212 samples were
grown by the traveling solvent floating-zone method, and include
underdoped samples ($T_c = 74 $ K), optimally doped samples ($T_c =
91 $ K), and overdoped samples ($T_c = 74 $ K). Hereafter, we will
refer to the samples with the usual ``T$_c$ notation'' such as UD74,
OP91, and OD74, where UD, OP, and OD mean under-doping,
optimal doping and overdoping, respectively. All samples were
cleaved \emph{in situ} and measured in an ultra-high vacuum with a base
pressure of better than 4 $\times$ 10$^{-11}$ mbar. For Na-CCOC samples
measurements were made at 10 K, while for Bi2212 samples
measurements were made at 200 K for UD74 and OP91 samples, and at 100 K
for OD74 samples\@.  It is worth noting that recent advances
\cite{KyungHee2006,Hanaguri2007} in high-pressure sample growth
resulted in the accessibility of the optimally doped and overdoped
samples of the Na-CCOC samples, for which no ARPES data have been
reported previously to our knowledge.

Figure 1 shows ARPES intensity maps at $E_F$ for Na-CCOC samples.  While
these data were taken at 10 K in the superconducting phase, the data will
be discussed here only in terms of the Fermi arc \cite{Norman1998} and
the pseudo-gap, with the more elusive and weaker feature of the
superconductivity left for a possible future study
\cite{KMShen2005,Hanaguri2007,gap}.

This figure clearly shows that the high-intensity region of the map
(``Fermi arc'' \cite{Norman1998}) increases its length as a function
of doping, as expected.  From our data, we define, operationally,
the ``Fermi surface'' as the contour determined by the momentum distribution
curve (MDC) peak positions (circle symbols). These Fermi surfaces in Figs. 1(a), and 1(b) are in
good agreement with those published previously on the same compounds
with similar doping values \cite{KMShen2005}.

Figure 2(a) shows the energy distribution curves (EDCs) near the
anti-nodal point, as a function of doping.  It is seen that these EDCs are
characterized by a large pseudo-gap, with the weight at $E_F$ strongly
suppressed.  This is quite reminiscent of a large pseudo-gap and the lack
of a coherent peak, as reported by STS \cite{Hanaguri2007}.

Figure 2(c) shows the MDCs for a cut
at the anti-nodal point for the UD21 sample, and Fig 2(d) for the
OP27 sample. In Figs. 1(c) and 2(d), MDCs are fit with two Lorentzian
curves (red dashed lines) and the positions of the two Lorentzian peaks are
marked with triangle symbols.  In agreement with a previous
work \cite{KMShen2005}, the MDCs for the underdoped sample
do not show appreciable dispersion up to a binding energy of $\sim 50$
meV\@.  However, the MDCs for the optimally doped sample show clear
dispersions at a binding energy of $\agt 30$ meV\@.  This is consistent
with the known qualitative behavior of the pseudo-gap as a function
of doping \cite{Damascelli}.

The Fermi surface as defined in Fig. 1 becomes flat as it crosses
the Brillouin zone boundary.  Thus the Fermi surface near the
anti-nodal point is approximately nested, as observed previously
\cite{KMShen2005}, and the nesting vector can be determined from our
Fermi surfaces.  This is shown in Fig. 3(a), which summarizes all
of our Fermi surfaces of Fig. 1.

\begin{figure}[t]
\setlength{\abovecaptionskip}{0pt}
\setlength{\belowcaptionskip}{0pt}
\begin{center}
\includegraphics[width=0.85\columnwidth]{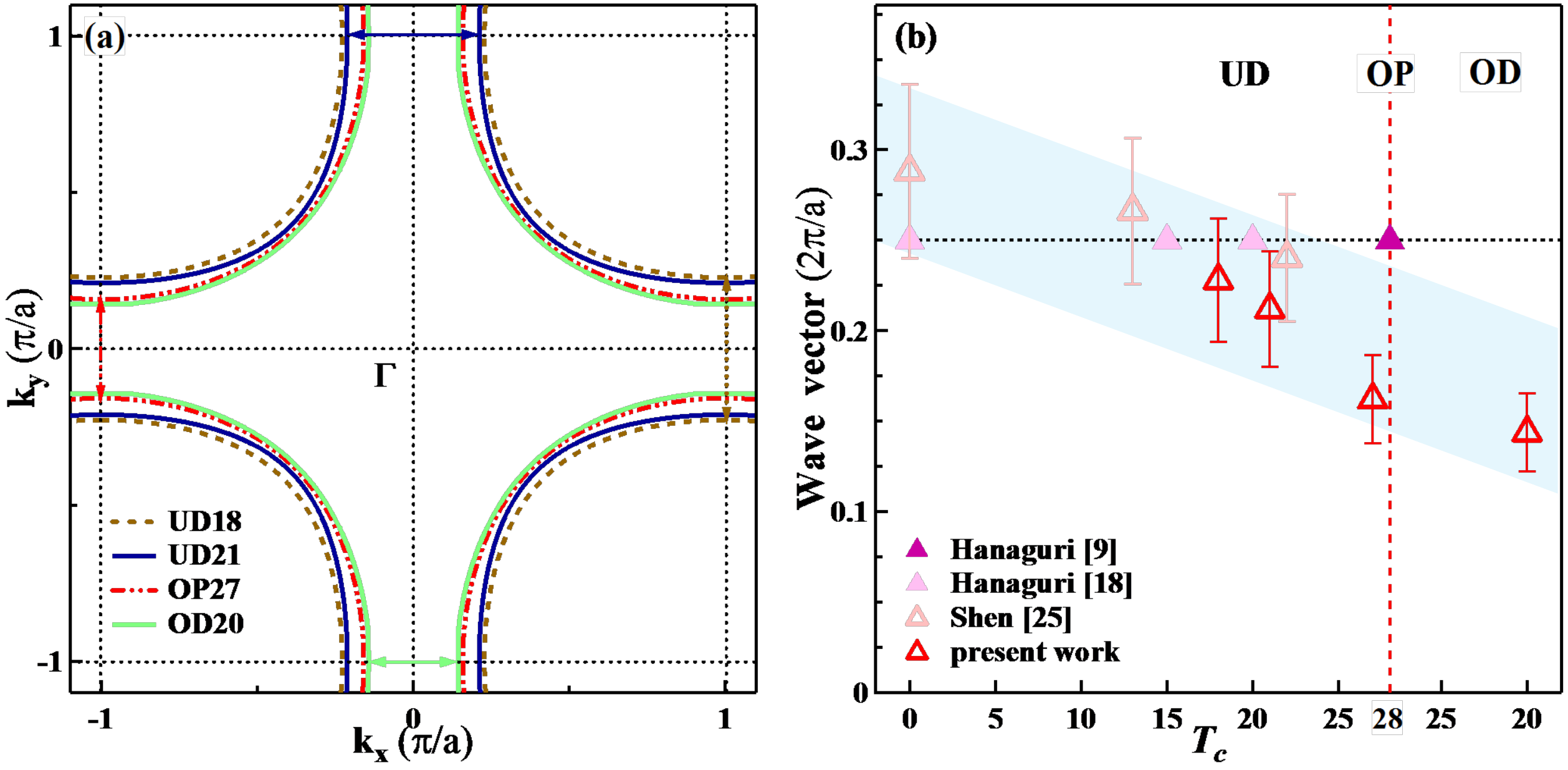}
\end{center}
\caption{(Color online) (a) ``Fermi surfaces'' as a function of doping for
Na-CCOC, corresponding to the data of Fig. 1. The arrows indicate the nesting vectors. (b)
FS nesting vectors (open symbols) and STS wave vectors (filled symbols).
The vertical dashed line (red) indicates the position of \emph{T$_{c,max}$}.}
\end{figure}

Figure 3(b) summarizes our nesting wave vectors, along with the wave
vectors corresponding to the checkerboard patterns of STS
\cite{Hanaguri2004,Hanaguri2007}.  Also, included in the figure are the
nesting wave vectors determined for underdoped samples of Na-CCOC by Shen
\emph{et al}.\@ \cite{KMShen2005}.  First of all, it can be easily noted that
our data agree well with Shen \emph{et al}.'s data for the overlapping
doping region.  Second, our data, now extended to optimal doping and
overdoping, clearly indicate that the nesting vector is a decreasing
function of doping.  Third, in hindsight, the data by Shen \emph{et al}.\@
also show the same trend.  Fourth, for the optimal doping, the FS nesting
would imply a CDW periodicity that is too large ($\agt 5.5 a$) as compared
with the checkerboard periodicity ($4 a$).

\begin{figure*}[floatfix]
\begin{center}
\includegraphics[width=1.7\columnwidth,angle=0]{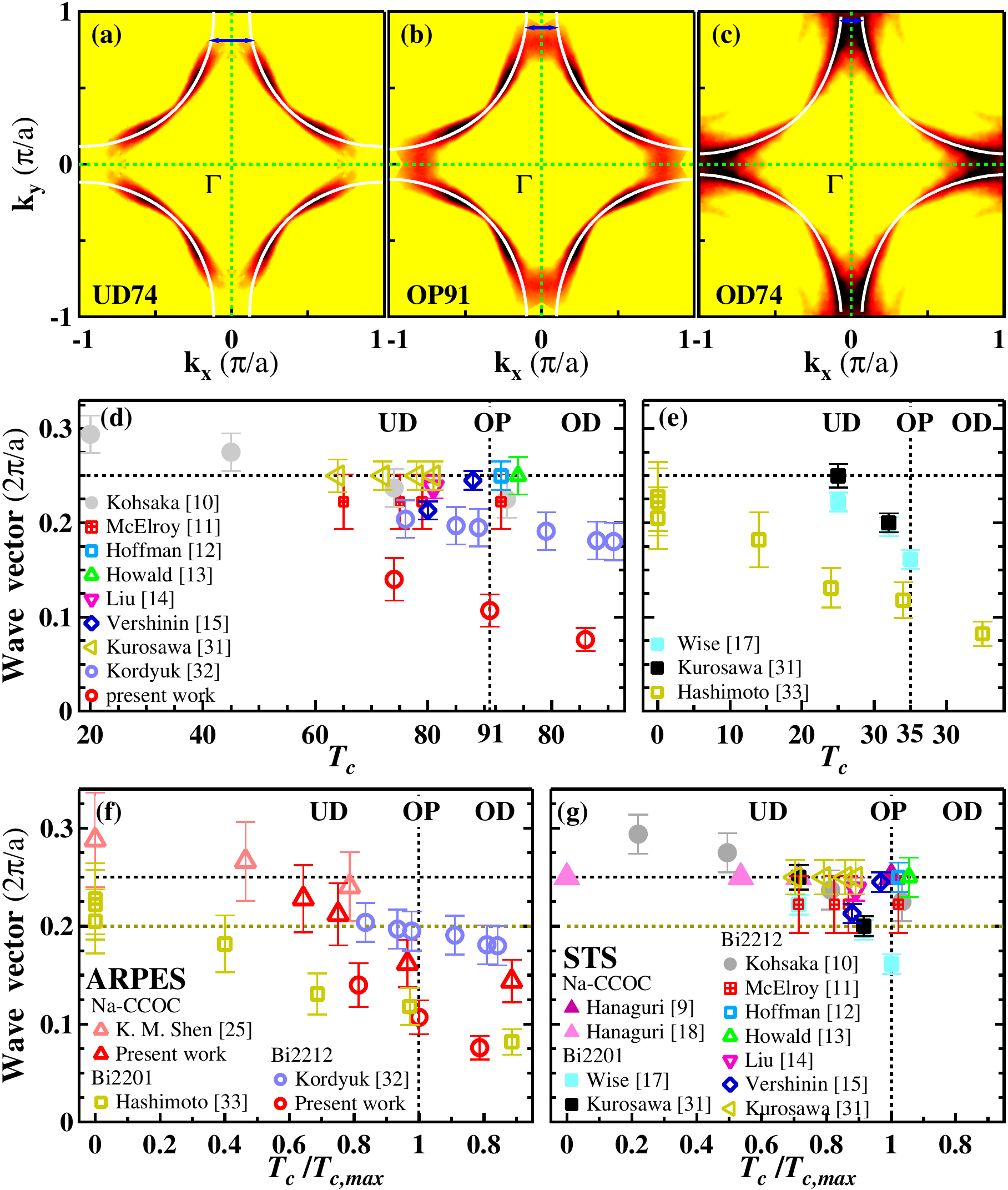}
\end{center}
\caption{(Color online) (a)-(c) Normal-state Fermi energy intensity maps of Bi2212.  White lines show fits to the Fermi surface of the tight-binding band theory \cite{Markiewicz2005}.
(d) The FS nesting vectors and the STS charge
order wave vectors for Bi2212, from this work and from the literature
\cite{Kohsaka2008, Hoffman2002A, Howald2003A, McElroy2005, Vershinin2004,
YHLiu2007, Kurosawa2010, Kordyuk2002}.  (e) Similar plot for Bi2201
samples.  All data are from the literature \cite{Hashimaoto2008, Wise2008,
Kurosawa2010}.  The formats of (d) or (e) are similar to those of Fig.3(b).
(f), (g) Summary of all ARPES nesting vectors (f) and STS charge order wave
vectors (g), showing their overall contrasting behaviors.}
\end{figure*}

Previously \cite{KMShen2005} the agreement between the ARPES nesting
vector and the checkerboard wave vector was interpreted to be very
important.  Given our findings above, however, there emerges another
possibility; The nesting wave vector and the checkerboard periodicity are
independent, although they might coincide by accident.  This is suggested
by the striking disagreement found in the doping dependence and a
similarly striking disagreement in the value at optimal doping, although
the second can be considered as a mere consequence of the first.

In order to help determine which case is more likely, we now discuss
Bi2212 superconductors and Bi2201 superconductors.  Fig. 4(a)-(c) show
ARPES intensity maps for Bi2212 samples.  Fig. 4(d) summarizes the
comparison of STS data and ARPES data for Bi2212.  The main feature is
that the Fermi-surface nesting vector for Bi2212 is always too small in
comparison with the checkerboard wave vectors.  Our experimental data
show the behavior of the anti-bonding band dispersion, while the data of
Kordyuk {\em et al\@.} \cite{Kordyuk2002}, included in Fig. (d), emphasize the
behavior of the bonding band dispersion, consistent with the well-known
empirical band structure of Bi2212 \cite{Markiewicz2005}.  The STS data
on Bi2212 suggest that the checkerboard periodicity is 3.5 - 4.5 lattice
constants, while the FS nesting would predict $\approx$ 5, or much
greater, lattice constants.  Clearly, Fig. 4(d) points out the difficulty
of reconciling the ARPES data with the STS data in the presence of
bilayer splitting, a point that has been already raised in a different
context \cite{Markiewicz-PRB-2004}.

Figure 4(e) shows the comparison for Bi2201 superconductors.  ARPES and
STS disagree greatly here also.  Furthermore, the disagreement becomes
{\em greater} as doping decreases, i.e.,\ as the pseudo-gap physics
becomes more important.  A related observation is as follows: While the doping
dependence of the STS data is in qualitative agreement with that of the
ARPES data, as pointed out previously \cite{Wise2008}, the slope of the
STS data is actually $\approx 2.5$ times too large in magnitude, as Fig. 4(e) shows.

Finally, Fig. 4(f), and 4(g) present all the data of Fig. 3(b),4(d), and 4(e) in a
different way.  The wave vectors are plotted as a function of normalized
$T_c$'s.  Two main discrepancies between ARPES and STS are observed.
First, ARPES data show a universal slope, while STS data do not.  Second,
STS wave vectors show a narrow distribution, with most values belonging in the
range $[0.2,0.25]$, while ARPES data show a wide and uniform distribution.

Putting aside this summary of discrepancies between ARPES and STS, one might ask
``is there a way to understand the apparent complexity in the data of
Bi2212 [Fig. 4(d)]?'' The presence of bilayer splitting in the ARPES
data and the mixed trends of the doping dependence in the STS data, the
latter presumably being due to different conditions of STS experiments,
contribute to this apparent complexity.  Fig. 4(f), which shows good
agreement between the bonding (anti-bonding) band data for Bi2212 with
the data for Na-CCOC (Bi2201), might be viewed as suggesting a possible
picture where the kinds of discrepancies observed in Na-CCOC [Fig. 3(b)]
and Bi2201 [Fig. 4(e)] have a mixed presence in Bi2212, giving rise
to the observed complexity.

To sum up, a weak-coupling CDW (or SDW) scenario is highly unlikely.
Instead a strong-electron correlation is likely the driving force of the
charge order.  Note that this conclusion does not rule out the possibility
of a cooperative or secondary effect of the FS nesting or a similar
momentum-space mechanism \cite{Kohsaka2008}.

Our claim is corroborated by more evidence.  First, if the FS nesting is
the driving force, then it is difficult to explain why the underdoping
region is preferred for the charge order.  According to a previous study
on Bi2212 \cite{Kordyuk2002}, the FS in the anti-nodal region becomes
very flat, and thus more highly nesting, at overdoping. Second, the FS
nesting scenario would require that the checkerboard pattern reverses the
contrast as the sample bias voltage changes sign \cite{Mallet1999}.  The
STS data \cite{Hanaguri2007} show quite the opposite: The contrast is
nonreversal.

The consideration of the so-called ``Luttinger sum rule'' \cite{Rice}
provides general additional support.  The sum rule concerns $x$, the hole
doping per unit cell.  In a strong correlation scenario, the sum rule is
satisfied only by the combination of the Luttinger surface (LS) and the FS.
We follow the theory of Ref.\@ \cite{Rice}, and take the LS to be the
anti-ferromagnetic zone boundary (AFM-ZB) and the FS to be approximately
the geometry formed by the Fermi arc and the AFM-ZB.  As indicated in the
inset of Fig. 1(a), in this scenario, the FS accounts for only $x$.  In a
weak-correlation scenario, the FS is now large, and accounts for $1+x$, as
indicated in Fig. 1(d).  Assuming the strong-correlation scenario, then,
we obtain $x = $ 0.11, 0.16, 0.21, and 0.23, respectively from UD18 through
OD20.  In the weak-correlation scenario, $x = $ 0.00, 0.06, 0.15, and 0.19.
So, while the latter scenario may be appropriate for optimal or over
doping, it fails completely at underdoping.  Our sum-rule analysis here
agrees well with a similar analysis based on the STS data of Bi2212
\cite{Kohsaka2008}, where a connection between the modified FS in the
strong correlation scenario and charge order was noted, as already discussed above,
and also with a recent ARPES work on Bi2212
\cite{Johnson-Fermi-Arc}.

To conclude, the electron correlation is of {\em primary} importance for
the charge order and the pseudo-gap in cuprates.  In
electron-correlation-based scenarios, the checkerboard periodicity has been
shown to be doping independent \cite{Seibold2007}, as in Na-CCOC, or an argument
based on the domain size has been given \cite{Anderson2004} to explain the
small doping dependence seen in Bi2212.  In closing, we note that similar
strong-correlation-based charge order scenarios have been proposed recently
for quasi-one-dimensional cuprates \cite{Rusydi2010} and La-based
superconductors \cite{Berg-FKT2009}, motivating further work.

G.H.G. acknowledges helpful discussions with S. A. Kivelson and D.-H. Lee.
G.H.G. and J.Q.M thank K. M. Shen for helpful discussions.
The work at UCSC was supported partially by a COR FRG grant.
Portions of this research were carried out at the Stanford Synchrotron Radiation Lightsource (SSRL), a Directorate of SLAC National Accelerator Laboratory and an Office of Science User Facility operated for the U.S. Department of Energy Office of Science by Stanford University.
The Advanced Light Source (ALS) is supported by the Director, Office of Science, Office of Basic Energy Sciences, of the U.S. Department of Energy under Contract No. DE-AC02-05CH11231.
The work at
the BNL was supported by DOE under Contract No. DE-AC02-98CH10886.


\begin{thebibliography}{99}
\bibitem{gap}{A possible feature in our data to be associated with the
    superconductivity is a few meV shift of the leading edge for the
    spectra within the Fermi arc. Given the energy window used in our
    experiment, this is a minor feature that needs further investigation.}
\end{thebibliography}
\end{document}